\documentstyle[12pt,aasms4]{article}

\begin{document}

\title{SELF-REGULATED ACCRETION DISKS}
\author{G. BERTIN}
\affil{Scuola Normale Superiore, Piazza dei Cavalieri 7, I 56126 Pisa,
Italy}
\authoremail{bertin@sns.it}
\slugcomment{Accepted for publication in The Astrophysical Journal Letters}

\begin{abstract}
We consider a class of fully self-gravitating accretion disks, for
which efficient cooling mechanisms are assumed to maintain the disk close to
the margin of Jeans instability. For such self-regulated disks the equations
become very simple in the outer regions, where the angular momentum convective
transport approximately balances the viscous transport. These latter equations
are shown to lead naturally to a self-similar solution with flat rotation 
curve, with circular velocity proportional to $(\dot{M})^{1/3}$ and 
essentially fixed opening angle.
\end{abstract}


\section{Introduction}

Studies of accretion disks have generally been motivated by
astrophysical phenomena in environments, such as X-ray binaries, 
where the presence of a
massive central object is established and dominates the scene. Under these
circumstances, the gravitational field in the accretion disk is basically
Keplerian and the self-gravity of the disk is expected to play a secondary role
(e.g., see Pringle 1981). On the other hand, there are regimes where the role
of self-gravity is important, so that the standard models
require modification. One step in this direction is
taken by considering the effects of self-gravity on the vertical structure of
the disk (Paczy\'{n}ski 1978) or by focusing on the impact of self-gravity on 
waves
and on transport processes (e.g., Lin \& Pringle 1987; Adams et al. 1989). In
particular, the role of Jeans instability in the disk has sometimes  been taken
into account by modifying the viscosity prescription (see Eqs.(14) and (16)
of the paper by Lin \& Pringle 1990).
In any case, much like in certain studies of planetary rings, it
is still customary to retain the assumption that the gravitational field that
sets the angular velocity of the disk is
basically Keplerian and spherically symmetric, even though there are
indications that under certain circumstances (e.g., for protostellar disks; see
Lin \& Pringle 1990; Drimmel 1996; and references therein) 
the disk mass may be significant.

In contrast with the case of planetary rings, in galaxy disks self-gravity
plays an additional non-trivial role, which is the shaping of the underlying
rotation curve (e.g., see van Albada \& Sancisi 1986). By analogy, one may
imagine conditions, for protostars or protogalaxies, where the disk forms {\it
without} a prominent  central object, and accretion initially proceeds mostly
under the action  of gravity within the disk
alone. These are precisely the conditions that this paper is devoted to.

The construction of models of accretion disks is generally based on the
specification of the relevant radiation processes associated with the disk
material, which gives a direct connection with the available observational
data. Unfortunately, since the data give only indirect clues on the microscopic
state of the disk, one has to resort to highly idealized models with a number
of free parameters. For the energy transport, viscous heating is determined from
the equations of hydrodynamics and radiation cooling is calculated under a
variety of assumptions on the disk opacity and on the radiation mechanisms
involved. The models are often based on a polytropic equation of state. In
turn, the amount of viscosity that is demanded by the observed phenomena
explained in terms of accretion disks is too large, well beyond that estimated
from the classical expressions for the viscosity coefficients derived from gas
dynamics or plasma physics. Then one is led to imagine that in the real
accretion disks collective phenomena are at work, producing ``anomalous
transport". This is not too surprising, since the inadequacy of the classical
transport  theory is well known in the context of laboratory and space plasmas
(e.g., see Coppi 1980). The traditional way to bypass such a stumbling block 
is to give a physically intuitive prescription for the
viscosity, where our ignorance of the underlying physical mechanisms is hidden
in a dimensionless parameter called $\alpha$ (Shakura \& Sunyaev 1973). This
also opens the way to a positive confrontation with the observations. It is
then  hoped that an examination of the available physical mechanisms (e.g., see
Balbus \& Hawley 1991) may eventually identify the origin of the large
viscosity required. 
 
While the momentum transport equations
are drastically simplified by means of the $\alpha$-prescription, in general the
energy equations are kept in their ideal form and followed in detail down to
their full consequences (e.g., see the equations describing advection-dominated
accretion; Popham \& Narayan 1991; 
Narayan \& Yi 1994).
One may really wonder whether a situation such as that of a protogalaxy or of a
protostar, where the cold infalling material may include gas clouds of
different sizes and different internal temperatures (in atomic or molecular
form), dust, and particulate
objects (as in planetary rings), should be described by a one-component
polytropic equation of state or whether, instead, such a complex collective
system should be modeled by a completely different approach, more in line with
the spirit that has led to the $\alpha$-prescription for the viscosity.

The purpose of this paper is to explore the consequences of one such an
alternate approach. We study a class of accretion disks where self-gravity
plays a full role and the energy equation is replaced by a self-regulation
prescription associated with the onset of Jeans instability
in the disk. A complete set of equations is thus proposed for the description
of a class of steady-state models. Surprisingly, the set of equations
immediately leads to a simple self-similar solution, with the disk density
inversely proportional to the radial distance and with flat rotation curve. The
opening angle set by the ratio $h/r$ is essentially fixed, while all the other
properties of the disk depend only on the value of $\alpha$ and on the value of
the accretion rate $\dot{M}$.

\section{Basic Equations for Self-Regulated Disks}

We consider a steady-state axisymmetric viscous inflow of a thin disk, with
constant mass ($\dot{M}$) and angular momentum ($\dot{J}$) accretion rates

\begin{equation}
\dot{M} = - 2 \pi r \sigma u~~,
\label{mass}
\end{equation}

\begin{equation}
\dot{J} = \dot{M} r^2 \Omega + 2 \pi \nu \sigma r^3 \frac{d\Omega}{dr}~~.
\label{angmom}
\end{equation}

\noindent Here $r$ is the radial coordinate in the plane of the disk, $\sigma$
is the disk density, $u$ is the radial velocity (so that an inflow, $u < 0$,
corresponds to a positive value of $\dot{M}$), $\Omega$ is the local angular
velocity of the disk, and $\nu$ is the relevant viscosity coefficient. Note
that for the common situation of a negative shear, $d\Omega/dr < 0$, the viscous
contribution to the angular momentum transport is negative, i.e. in the outward
direction.

For a cool, slowly accreting disk the radial momentum balance equation requires

\begin{equation}
\Omega^2 \sim 
\frac{1}{r}\frac{d \Phi_{\sigma}}{d r} + \frac{G M_{\star}}{r^3}~~,
\label{radmom}
\end{equation}

\noindent where $M_{\star}$ is the mass of the central object and
$\Phi_{\sigma}$ is the contribution to the gravitational field provided by the
mass of the disk

\begin{equation}
\Phi_{\sigma}(r) = - 2 \pi G \int^{\infty}_0 K^{(0)}(r, r') \sigma(r') d r'~~,
\label{poisson}
\end{equation}

\noindent with $K^{(0)}(r, r') = (1/\pi) \sqrt{r' \zeta/ r} K(\zeta)$. Here
$\zeta \equiv 4 r r' /(r + r')^2$ and $K(\zeta)$ is a complete elliptic
integral of the first kind. [Corrections to Eq.(\ref{radmom}) related to the
disk pressure gradient and to the convective gradient $u (du/dr)$ can be easily
incorporated, but do not change the following discussion.]

The viscosity in the disk is taken according to the Shakura \& Sunyaev (1973)
prescription

\begin{equation}
\nu = \alpha c h~~,
\label{ss}
\end{equation}

\noindent where the half-thickness $h$ is related to the effective thermal
speed $c$ following the hydrostatic requirement of a self-gravitating disk

\begin{equation}
h = \frac{c^2}{\pi G \sigma}~~.
\label{vert}
\end{equation}                                                    

\noindent Note that inserting (\ref{vert}) into (\ref{ss}) yields the simple
relation $\nu \sigma = (\alpha/\pi G) c^3$, which can be inserted further in
Eq.(\ref{angmom}) to give

\begin{equation}
G \dot{J} = G \dot{M} r^2 \Omega + 2 \alpha c^3 r^3 \frac{d \Omega}{d r}~~.
\label{angmom1}
\end{equation}

Even with $\alpha$ treated as an ``assigned parameter", the equations written
so far form an incomplete set. As mentioned in the Introduction, the set is
usually closed by means of an energy equation, or by an equation balancing
viscous heating and radiation losses. Here we explore the possibility of closing
the set by a self-regulation prescription (see Bertin 1991) related to the Jeans
instability of the disk. The physical argument is that, in the presence of Jeans
instability, on the fast dynamical timescale the disk is bound to heat up until
it becomes marginally stable; if the disk is warmer to begin with, it is assumed
that dissipation, e.g. in the form of inelastic collisions of the fluid elements
of the disk, make the effective thermal speed $c$ quickly drop to the marginal
value. The condition of marginal stability is of the form

\begin{equation}
\frac{c \kappa}{\pi G \sigma} = \bar{Q}~~, 
\label{jeans}
\end{equation}

\noindent with $\bar{Q} \approx 1$. Here $\kappa$ is the epicyclic frequency.

Massive self-gravitating disks with $c\kappa/\pi G \sigma$ close to unity are 
known to be subject to violent spiral instabilities (e.g., see Bonnell 1994,
Laughlin \& Bodenheimer 1994; for the case of galaxy disks, see Bertin et al.
1989). Therefore, disks of the kind considered in this paper may well be in a
situation where the gravitational torques associated with non-axisymmetric
instabilities are able to contribute a significant amount of ``anomalous
viscosity". In this case, there might be no need for other mechanisms (such as
those, often invoked, related to magnetic fields) to drive the overall accretion
process in the disk. The hope is that the axisymmetric model
described here, while obviously oversimplifying the actual state of the disk, is
able to capture the basic dynamical properties of a fully self-gravitating
accretion process. 

In view of the above remarks, it would be desirable to add one further relation, 
between $\alpha$ and $\bar{Q}$, in order to properly incorporate the
gravitational instability contributions to the viscosity prescribed by 
Eq.(\ref{ss}). Previous
studies (Lin \& Pringle 1987, 1990; Laughlin \& Bodenheimer 1994) have focused
on the role of self-gravity on viscous transport and indicate that there is some
merit in an axisymmetric model with modified viscosity. Here we take a
drastically different approach in that we retain the standard functional
dependence of the viscosity (see Eq.(\ref{ss})) and we use  the marginal
condition (\ref{jeans}) to replace the energy equations altogether. As to the
use of the $\alpha$-prescription, it has been noted (see Laughlin \&
R\'{o}\.{z}yczka 1996, who based their analysis on hydrodynamical simulations
of a polytropic thin disk) that $\alpha$ should be allowed to vary with radius
and that the prescription may be inadequate {\it a priori},  since it is
inherently local, while the self-excited non-axisymmetric modes involved are
global. In the absence of a convincing global model, we have decided to explore
first the consequences of the simple constant-$\alpha$ case. {\it A posteriori},
at least for the {\it self-similar solution} described in the next section,
this simplifying assumption is quite natural and may well turn out to be
dynamically consistent.  

For specified values of $\alpha$ and $\bar{Q}$, equations (\ref{radmom}),
(\ref{angmom1}), and (\ref{jeans}) thus form a complete set (we take
$\Phi_{\sigma}$ defined by (\ref{poisson})) with solutions determined by three physical
parameters, i.e. $M_{\star}$, $\dot{J}$, and $\dot{M}$. One may  argue that at
large radii the effects of $M_{\star}$ and $\dot{J}$ eventually  become
unimportant and thus deal with simple (approximate) relations that are 
appropriate for the outer disk. Thus we are led to the following reference case.

\section{Flat Rotation Curve}

In this paper we focus on the special case where $M_{\star} = 0$ and
$\dot{J} = 0$. The basic equations are then reduced to

\begin{equation}
\Omega^2 = \frac{1}{r} \frac{d \Phi_{\sigma}}{d r}~~,
\label{radmom2}
\end{equation}

\begin{equation}
2 \alpha c^3 = G \dot{M} \left|\frac{d \ln{\Omega}}{d \ln{r}}\right|^{-1}~~,
\label{angmom2}
\end{equation} 

\begin{equation}
\left(\frac{c}{\pi G \sigma}\right) 2 \Omega 
\left(1 + \frac{1}{2} \frac{d \ln{\Omega}}{d \ln{r}}\right)^{1/2} = \bar{Q}~~,
\label{jeans2}
\end{equation}

\noindent where, we recall, we have in mind $\dot{M} > 0$, 
$0 < \alpha \lesssim 1$, and
$\bar{Q} \approx 1$. The equivalent thermal speed $c$ can be eliminated to give a
relation between density and angular velocity

\begin{equation}
\pi G \sigma = \frac{2 \Omega}{\bar{Q}}\left(1 + 
\frac{1}{2} \frac{d \ln{\Omega}}{d \ln{r}}\right)^{1/2} 
\left(\frac{G \dot{M}}{2 \alpha}\right)^{1/3} 
\left|\frac{d \ln{\Omega}}{d \ln{r}}\right|^{- 1/3}~~.
\label{jeans3}
\end{equation}

\noindent Therefore an obvious natural solution is provided by the
self-similar mass distribution characterized by

\begin{equation}
2 \pi G \sigma r = r^2 \Omega^2 = V^2 = const~~,
\label{sigss}
\end{equation}

\noindent which solves both (\ref{jeans3}) and the self-consistency condition
(\ref{radmom2}). Thus we find

\begin{equation}
c = \left(\frac{G \dot{M}}{2 \alpha}\right)^{1/3} \approx 
\left(\frac{27 \dot{M}}{4 \pi \alpha}\right)^{1/3}~~10~km/sec~~,
\label{css}
\end{equation}

\begin{equation}
V = \frac{2\sqrt{2}}{\bar{Q}} c~~,
\label{vss}
\end{equation}

\begin{equation}
u = - \frac{\alpha \bar{Q}^2}{4} c~~,
\label{uss}
\end{equation}

\begin{equation}
\frac{h}{r} = \frac{\bar{Q}^2}{4}~~,
\label{hss}
\end{equation}

\begin{equation}
\nu \sigma = \frac{\dot{M}}{2 \pi}~~.
\label{nuss}
\end{equation}

The self-consistent solution provided above is a generalization of the
self-similar solution for self-gravitating disks (Mestel 1963) to the case of
viscous disks. It has infinite mass and a singularity at $r = 0$, which is 
even more evident here in the viscous case, since a constant accretion rate 
implies a constant accumulation of mass at the origin. Since the value of 
$\bar{Q}$ is well constrained by dynamics, the
final solution depends only on the value of the parameter $\alpha$ and on the
accretion rate $\dot{M}$. Note that in the last expression of Eq.(\ref{css}) 
$\dot{M}$ is to be given in units of $M_{\odot}/yr$. One interesting feature is
that the opening angle of the disk is fixed ($\approx 14$ degrees for 
$\bar{Q} = \sqrt{2}/2$), 
set by Eq.(\ref{hss}), independent of $\alpha$ and of the mass accretion rate. 

\section{Discussion}

Here below we list a few points that we
plan to investigate in a following article. A resolution of these issues should
be able to bring down the singular self-similar disk to more realistic
conditions. 

(i) Within the assumptions of the previous Section, we note that values of
$\bar{Q}$ significantly larger than unity would be undesired, because they 
go against the thin disk approximation (cf. Eqs.(\ref{vss}) and (\ref{hss})).
However, thickness itself helps to cure the problem; it is possible to show 
that the effects of finite thickness reduce $\bar{Q}$ from its reference value 
of unity down to $\approx \sqrt{2}/2$. On the other hand, a more thorough 
investigation of the plausible range of $\bar{Q}$ in a self-regulated disk is
desired.

(ii) The singularity at $r = 0$ should be removed, especially since, for a large
class of systems, the central region is the astrophysically interesting part of
the disk. Relaxing the assumptions of $\dot{J} = 0$ and $M_{\star} = 0$ will
certainly open the way to models that have a better physical justification. In
particular, a smooth, self-regulated process is better justified outside 
the extreme regime where all the mass is in the disk. Note
that, for given values of $M_{\star}$, $\dot{M}$, and $\alpha$, the  rotation
curve will be modified at radii smaller than or of the order of  $r_S \equiv (G
M_{\star}/4)(G \dot{M}/2 \alpha)^{-2/3}$.  Furthermore, independently of the
mass of the central object, we expect a  change in the properties of the
solution on a scale defined by  $r_J \equiv (|\dot{J}|/4 \dot{M}) (G \dot{M}/2
\alpha)^{- 1/3}$. A number of regimes are available depending on the relative
size of the two lengthscales  $r_S$ and $r_J$. We recall that the equations
studied in the previous Section are approximately correct at large radii, i.e.
at  $r > r_S$, $r > r_J$.

(iii) For specific astrophysical systems, one should try to set also a proper
boundary condition at large radii, and check to what extent an approximately
self-similar solution can be brought to match the physical conditions there.
For example, in the context of binary stars where one star supplies mass to the
other, while there may be a
significant radial range available for the equations of the previous Sections
to be approximately correct, $r$ should not exceed the relevant Roche radius.

(iv) Especially from the expression for the thermal speed (\ref{css}), it is
clear that, unless unrealistic values for the accretion rate are considered,
the self-similar solution calculated above has very little to say about X-ray
emitting astrophysical disks. In view of this difficulty, it is important to
explore the possibility that in some radial range, presumably close to the
center, the self-regulation constraint may break down or, rather, be replaced
or overruled by a different energy balance equation. In turn, this would open
the possibility of hotter disks and, in any case, of disks for which the
``rigid" structure of the self-similar solution is relieved. [In spiral
galaxies, the self-regulation process is expected to take place in the outer
disk, but generally to break down at small radii, where $c$ increases
significantly above the marginal value in most realistic models (see Bertin et
al. 1989).]

While we were working on the natural follow-up extensions (i) - (iv) of the
self-similar solution found in Section 3, we have noted that self-similar,
self-gravitating disks have been discussed, very recently, by Mineshige \&
Umemura (1996). In spite of the apparent similarity of their solution to ours
(they also give a prominent role to self-gravity both in the vertical and in
the horizontal directions and they also find Mestel solutions for the mass
distribution), there are significant differences between their work and the
present article, especially at the physical level. In
particular, in their article: (a) The starting point is an entropy transport
equation (instead of our condition (\ref{jeans})); (b) The self-similar
solution is sought from the beginning, then their finding that for a
self-gravitating disks the only power-law solutions are those with $r \Omega =
const$ is not surprising (while in our Section 3 the flat rotation curve
turns out, {\it a posteriori}, to be the natural solution for a steady-state
self-regulated disk); (c) Isothermality is forced by requiring the disk density
to follow the behavior of the Mestel disk, but is inconsistent
with possible cooling scenarios (either optically thin free-free emission or
black-body radiation) that determine the entropy transport equation (our
solution is instead internally consistent); (d) The opening angle of the disk 
(i.e., 
$h/r$) is a free parameter (while it is fixed  in the present paper; see
Eq.(\ref{hss})); (e) The realization that there is a simple, dimensionally
interesting connection between the thermal speed $c$ in the self-similar disk
and the accretion rate (see our Eq.(\ref{css})) is apparently missed.

In conclusion, based on the assumption that the self-gravity of the
disk plays a dominant role, we have identified here a simple analytical
solution for self-regulated, steady-state accretion disks that appears to be
an interesting starting point for the construction of a new class of accretion
disks. Quite unexpectedly, we have shown that by promoting self-gravity to 
its full role one is naturally led to flat rotation curves.  

\acknowledgments

I would like to thank Bruno Coppi for several interesting discussions. Much of
this work has been carried out during a visit at MIT and completed a few
months later, during a visit at the STScI. The two institutions are thanked for
their hospitality. This work has been partially supported by MURST and ASI of
Italy.

\end{document}